\documentstyle[12pt]{article}

\begin{document}
\title{Dynamical symmetries and the Ermakov invariant}
\author{
F.~Haas\footnote{ferhaas@lncc.br \hskip 10pt
$^\dag$goedert@exatas.unisinos.br}\\
Laborat\'orio Nacional de Computa\c{c}\~ao Cient\'{\i}fica - LNCC \\
Av. Get\'ulio Vargas, 333\\
25651-07 Petr\'opolis, RJ, Brazil\\
J.~Goedert$^\dag$\\
Centro de Ci\^encias Exatas e Tecnol\'ogicas - UNISINOS\\
Av. Unisinos, 950\\
93022-000 S\~ao Leopoldo, RS, Brazil}
\date{\strut}
\maketitle

\begin{abstract}
Ermakov systems possessing Noether point symmetry are identified
among the Ermakov systems that derive from a Lagrangian formalism
and, the Ermakov invariant is shown to result from an associated
symmetry of dynamical character.  The Ermakov invariant and the
associated Noether invariant, are sufficient to reduce these systems
to quadratures.
\end{abstract}

\leftline{PACS number(s): 02.30.Hg, 02.90.+p, 03.20.+i}
\section{Introduction}

Constants of motion (invariants or first integrals) are of central
importance to study dynamical systems in general.
This has motivated the development of several procedures
by which constants of motion for dynamical systems (see \cite{Kaushal}
for a recent review)  can be constructed. Among these, the
application of Noether's theorem \cite{Sarlet,Haas1} deserves special
attention in view of the physical appealing of the method. Indeed,
Noether's theorem provides links between continuous symmetries of the
action functional and constants of motion of the system.  Among the
classical results obtained by application of Noether's theorem are
the conservation of energy, linear momentum and angular momentum. The
associated symmetries correspond to the invariance of the action
functional under time translation, space translation  and  space
rotations, respectively.

Recently, Ermakov systems \cite{Ermakov} were analyzed
\cite{RR2}--\cite{Kaushal2} via Noether's theorem.  Ermakov systems,
which are pairs of coupled, second--order ordinary differential
equations, always admit at least one constant of motion, the Ermakov
invariant.  The work developed in \cite{RR2}--\cite{Kaushal2}
provided a consistent explanation for both the equations of motion
and the invariant of Ermakov systems, as a result of an underlying
Noether symmetry.  The approach proposed, however, applied only to
Ermakov systems where one of the equations  decoupled from the other.
This limitation has motivated comments \cite{Kana} concerning the
difficulty of deriving the Ermakov invariant from symmetry
considerations only.  A closer examination shows that references
\cite{RR2}--\cite{Kaushal2} assumed that  one of the variables in the Ermakov
system plays the central role, the other being regarded as a mere
auxiliary variable.  This point of view naturally led to systems
where one of the equations (the so called auxiliary equation)
decoupled from the other, of a more important character. In the
present work, we  consider the Ermakov systems as essentially
coupled, with no dynamical variable playing any distinguished role.
For this, we propose for the Ermakov systems not a one--dimensional
but a two--dimensional Lagrangian formalism.  Using this Lagrangian
description, we search to construct coupled Ermakov systems for which
Noether's theorem applies. In this context, both the Ermakov and all
additional invariants - when they exist - should result from the
associated Noether symmetry which then provides a general framework
to define, classify and solve Ermakov systems.
We also remark that the use of a two-dimensional Lagrangian formalism
for Ermakov systems has been suggested in a recent work \cite{Simic}.

A central issue with Ermakov systems concerns their complete integrability.
Specificaly, we analyse the complete integrability of Ermakov systems
in the Liouville sense \cite{Arnold}, which, in turn, requires the existence
of an underlying Hamiltonian structure.
In fact, complete integrability requires, in
this cases, the existence of at least a second invariant.  The
existence of Ermakov systems with a second invariant was first
addressed by Goedert \cite{Goedert}, who obtained a class of Ermakov
systems possessing a second constant of motion.
For Hamiltonian Ermakov systems, Liouville's theorem \cite{Arnold}
warrants complete integrability if a second constant of motion exists
in involution with the Ermakov invariant. Notice also
that, strictly speaking, Liouville's theorem requires compact level
surfaces of the constants of motion.

The present work focus on the application of Noether's theorem to
Ermakov systems.  The Lagrangian formalism constitutes the usual
framework for the application of Noether's theorem \cite{Sarlet} (and
references therein).  We thus need Ermakov systems that may be
derived from a Lagrangian formalism.  This has already been dealt
with in the literature \cite{Cervero, Haas2}.  Cerver\'o and
Lejarreta \cite{Cervero} showed that conventional Hamiltonian Ermakov
systems are completely integrable. By conventional Ermakov systems,
we mean  Ermakov systems having a frequency function depending on
time only.  However, as pointed out previously  \cite{Reid}, a
frequency function depending also on the dynamical variables does not
destroy the main feature of Ermakov systems, namely the existence of
the Ermakov invariant. This has motivated a number of studies
\cite{Goedert}, \cite{Haas2, Saermark, Haas3} on Ermakov systems with
a generalized frequency function, depending on dynamical variables
besides time.  Following this trend, we do not restrict the frequency
functions dependence on time only. Nevertheless we will show that the
integrability of conventional Hamiltonian Ermakov systems results
from a Noether symmetry, a fact that was not clear at all in the
original reference \cite{Cervero}.
Finally, we remark that dynamical symmetries have been considered in
reference \cite{Athorne1} for Ermakov
systems of conventional type and not necessarily
having Lagrangian character.  Let us stress that in the present
work the interest is centered on Noether symmetries, which may be
assessed only when a variational description of the system is
available. Besides, we do not restrict ourselves to conventional
Ermakov systems.

The paper is organized as follows.  In section 2, we obtain a
class of Ermakov systems admitting a Lagrangian description,
directly from the class of Hamiltonian Ermakov systems derived in
reference \cite{Haas2}. In section 3 we show, by use of the converse
of Noether's theorem \cite{Sarlet}, that the Ermakov invariant is
the result of a Noether dynamical symmetry.  Within the class of
Lagrangian Ermakov systems, we also find the subclass of systems that
possess Noether point symmetry. The Lagrangian Ermakov systems with
Noether point symmetry, having a second constant of motion in
involution with the Ermakov invariant, are Liouville integrable.

\section{Lagrangian Ermakov systems}

Ermakov systems are generally \cite{Haas2} defined as the system of
equations
\begin{eqnarray}
\label{eq1} \ddot x + \omega^{2}x &=& \frac{1}{yx^2}\,f(y/x)
\,,\\ \label{eq2} \ddot y + \omega^{2}y &=& \frac{1}{xy^2}\,g(x/y) \,,
\end{eqnarray}
where $f$ and $g$ are arbitrary functions of the indicated
arguments and $\omega$ is an arbitrary function of time, the
dynamical variables $x$ and $y$, and their time derivatives.
For physical reasons, it is convenient to restrict the dependence to
$\omega = \omega(t,x,y,\dot{x},\dot{y})$.  For brevity, we call
$\omega$ the frequency function, even if it is not associated with
any characteristic frequency of the system.  We recall that, as shown
in \cite{Haas2}, only one arbitrary function of $y/x$ is sufficient
to specify a generic Ermakov system if the frequency function is
allowed to depend on the dynamical variables.  However, for easier
comparison with the results in the literature, we take
(\ref{eq1}--\ref{eq2}) as our definition of Ermakov systems.

The central property of Ermakov systems is their Ermakov
invariant,
\begin{equation} \label{eq3} I = \frac{1}{2}(x\dot{y} -
y\dot{x})^2 + \int^{y/x}f(\lambda)d\lambda +
\int^{x/y}g(\lambda)d\lambda \,.  \end{equation}
The Ermakov invariant is constant along the trajectories of
(\ref{eq1}--\ref{eq2}) independently of the detailed dependence of
$\omega$ on its arguments.  The Ermakov invariant is most useful  to
construct a nonlinear superposition law \cite{Reid} relating the
solutions of (\ref{eq1}) and (\ref{eq2}).  Here, however, we are
interested in Ermakov invariants as one ingredient for the
derivation of completely integrable Ermakov systems.

Given the variational formulation, we can use some more
specialized tools to investigate integrability, such as Noether's
theorem \cite{Sarlet}.  This fact makes it interesting to
consider not the general Ermakov systems (\ref{eq1}--\ref{eq2}), but
only their subclasses that admit a variational formulation. In
particular, we shall focus on Lagrangian descriptions, since it is on
configuration space that Noether theorem is usually formulated
\cite{Sarlet}.

We, therefore, pose the following question: under what conditions on
the functions $\omega$, $f$ and $g$ does (\ref{eq1}--\ref{eq2})
possess a variational description?  This question was partially
answered by Cerver\'o and Lejarreta \cite{Cervero}, who considered
Ermakov systems with usual frequency functions, and by Haas and
Goedert \cite{Haas2}, in the case of frequency functions depending on
dynamical variables. Both works adopted a variational description  of
the Hamiltonian type. Here, however, we are specially interested in
Lagrangian descriptions and their connection with Noether's theorem.
Therefore, we rewrite the results of reference \cite{Haas2} in terms
of a Lagrangian formalism.  Here this is the more direct approach, even
if the usual procedure is to derive the Hamiltonian formalism
from the Lagrangian description.

Using the results of \cite{Haas2} we may consider
(\ref{eq1}--\ref{eq2}) as obtained from a Lagrangian function that
represents the sum of a quadratic kinetic energy term and a potential energy
term provided that the frequency function is taken in the form
%
\begin{equation}
\label{eq13} \omega^2 = \frac{1}{R}\frac{\partial
\bar{V}}{\partial R}(R,t) + \frac{\sigma(\theta)}{R^4} \,,
\end{equation}
where $\sigma(\theta)$ is defined as
\begin{eqnarray}
\sigma(\theta) &=& \frac{(A\cos\theta +
B\sin\theta)}{\psi^{2}(\theta)\sin\theta\cos^{2}\theta}\,f(\tan\theta) +
\frac{(B\cos\theta +
C\sin\theta)}{\psi^{2}(\theta)\sin^{2}\theta\cos\theta}\,g(\cot\theta) \nonumber \\
&-& 2\kappa\left(\int^{\tan\theta}f(\lambda)d\lambda +
\int^{\cot\theta}g(\lambda)d\lambda\right)  \,,
\end{eqnarray}
$f$ and $g$ being kept totally arbitrary.
In (\ref{eq13}) $\bar{V}(R,t)$ is an arbitrary function, and,
$R$ and $\theta$ are defined by
\begin{equation}
\label{eq16} R^2 = Ax^2 + 2Bxy + Cy^2 \,, \qquad \theta =
\arctan(y/x) \,,
\end{equation}
with $A$, $B$, $C$ and $\kappa$ numerical constants satisfying
\begin{equation}
\label{eq10} \kappa \equiv AC - B^2 \neq 0 \,.
\end{equation}
Also, in (\ref{eq13}) we used the notation
\begin{equation}
\label{eq18}
\psi^{2}(\theta) = \left(A\cos^{2}\theta +
2B\sin\theta\cos\theta + C\sin^{2}\theta\right)^{-1}\,.
\end{equation}
As mentioned before, in general $\omega$ does not represent any
physical frequency of the system, but we keep the nomenclature for
convenience of notation. The associated Lagrangian function is then
\begin{equation}
\label{eq4} L = T - V \,,
\end{equation}
where the kinetic energy is
\begin{equation}
\label{eq5} T = \frac{1}{2}(A\dot{x}^2 +
2B\dot{x}\dot{y} + C\dot{y}^2) \,,
\end{equation}
and
\begin{equation}
\label{eq15} V = \bar{V}(R,t) +
\frac{\kappa}{R^2}\left(\int^{\tan\theta}f(\lambda)d\lambda +
\int^{\cot\theta}g(\lambda)d\lambda\right) \,.
\end{equation}
%
For the sake of completeness, we writte the corresponding
Hamiltonian,
\begin{equation}
\label{HHH}
H = \frac{1}{2\kappa}(Cp_{x}^2 + 2Bp_{x}p_{y} + Ap_{y}^2) + V \,.
\end{equation}

An important point concerning Ermakov systems is their linearisation. As shown in
\cite{Athorne2}, Ermakov systems with frequencies depending only on time are always reducible to a linear second-order ordinary differential equation. Latter on
\cite{Haas4}, the same linearisation was shown to apply to a subclass of Ermakov systems with frequency functions depending on dynamical variables. In the final part of the section, we verify in what cases the Lagrangian Ermakov systems just described may be
linearisable following the procedure of references \cite{Athorne2, Haas4}. This procedure relies on the introduction of a new dependent variable $\varphi$ given by
\begin{equation}
\label{varphi}
\varphi = \alpha(t)/R \,,
\end{equation}
where $\alpha(t)$ is a function of time to be choosen conveniently. Also, we must use the polar angle $\theta$ as the new independent variable. The change of independent variable is acomplished by the relation
\begin{equation}
\label{thetap}
\dot\theta = h(\theta;I)/R^2 \,,
\end{equation}
which follows from the Ermakov invariant, where
\begin{equation}
\label{htheta}
h(\theta;I) = \frac{\sqrt{2}}{\psi^{2}(\theta)}\left(I - \int^{\tan\theta}f(\lambda)d\lambda + \int^{\cot\theta}g(\lambda)d\lambda\right)^{1/2} \,.
\end{equation}
Actually, in references \cite{Athorne2, Haas4} the variables used are $\alpha(t)/r$ and $\theta$, where $(r,\theta)$ are the conventional polar coordinates, but here the choice (\ref{varphi}-\ref{htheta}) can be shown to be more effective.

Using the Lagrangian equations of motion obtained from (\ref{eq4}) and the relations (\ref{varphi}-\ref{htheta}), we get
\begin{equation}
\label{quasi} h^{2}\frac{d^{2}\varphi}{d\theta^2} +
h\frac{\partial h}{\partial\theta}\frac{d\varphi}{d\theta} +
2\kappa\,I\varphi =
\alpha^{2}\left(\frac{\alpha\ddot\alpha}{\varphi^3} -
\frac{\partial\chi}{\partial\varphi}\right) \,,
\end{equation}
where $\chi$ is $\bar{V}$ expressed in terms of the new dependent variable,
\begin{equation}
\chi = \chi(\varphi,t) = \bar{V}(\alpha/\varphi, t) \,.
\end{equation}
The left-hand side of eq. (\ref{quasi}) is a linear form. Obviously, the complete equation will have a linear character if and only if the right-hand side of (\ref{quasi}) is a linear form, which imposes
\begin{equation}
\label{chi}
\chi = \bar{\chi}(t) + \frac{a\varphi^2}{\alpha^2}
- \frac{b\varphi}{\alpha^2} - \frac{\alpha\ddot\alpha}{2\varphi^2} \,,
\end{equation}
where $\bar{\chi}(t)$ is an arbitrary function of time which can be set to zero without loss of generality since it has no influence on the equations of motion. Moreover,
$a$, $b$ are arbitrary numerical constants. The resulting linearisation reads
\begin{equation}
\label{quasi2} h^{2}\frac{d^{2}\varphi}{d\theta^2} +
h\frac{\partial h}{\partial\theta}\frac{d\varphi}{d\theta} +
2(\kappa\,I + a)\varphi = b \,.
\end{equation}
The reconstruction of the solution of the original Ermakov system from the solution of (\ref{quasi2}) is made following the steps described in \cite{Athorne2}.

Here, the important point is that not all Lagrangian Ermakov systems are linearisable in terms of the proposed change of variables. Indeed, (\ref{chi}) imposes a particular  dependence of the originally arbitrary function $\bar{V}$, namely
\begin{equation}
\label{linearV}
\bar{V}(R,t) = \frac{a}{R^2} - \frac{b}{\alpha\,R} - \frac{\ddot\alpha\,R^2}{2\alpha} \,.
\end{equation}
For $\alpha =$ cte., the resulting equations of motion can be shown to be in the class of
Kepler-Ermakov systems \cite{Athorne5}. For general $\alpha(t)$, we have a generalized time-dependent Kepler problem \cite{Munier}.

\section{Noether symmetries}

Let us consider the transformations
\begin{eqnarray}
\label{eq24} \bar{x} &=& x + \varepsilon\,\eta_{1}(x,y,\dot{x},\dot{y},t) \,,
\nonumber\\
\bar{y} &=& y + \varepsilon\,\eta_{2}(x,y,\dot{x},\dot{y},t) \,, \\
\bar{t} &=& t + \varepsilon\,\tau(x,y,\dot{x},\dot{y},t) \,, \nonumber
\end{eqnarray}
where $\varepsilon$ is an infinitesimal parameter and $\eta_1$,
$\eta_2$ and $\tau$ are functions to be determined. As these
functions may depend on the velocities, we are dealing with
transformations of dynamical type.  Dynamical symmetries may also
contain a dependence of $\tau$, $\eta_1$ or $\eta_2$ on higher
derivatives, but, for simplicity, we restrict the treatment to
symmetries of form (\ref{eq24}).

To proceed we introduce the following more convenient notation,
\begin{equation}
{\bf q} = (x,y) \,,\qquad {\bf\eta} = (\eta_{1},\eta_{2}) \,,
\end{equation}
and consider the action functional
\begin{equation}
\label{eq25} S[{\bf q}] = \int_{t_0}^{t_1}L\,dt \,.
\end{equation}
Noether's theorem establishes a correspondence between continuous
transformations that leave the action functional invariant (up to
an additive numerical constant) and conservation laws. In particular,
Noether's symmetry criterion \cite{Sarlet} states that
\begin{equation}
\label{eq26} \tau\frac{\partial L}{\partial t} +
{\bf\eta}\cdot\frac{\partial L}{\partial{\bf q}} + (\dot{\bf\eta}
- \dot\tau\dot{\bf q})\cdot\frac{\partial L}{\partial\dot{\bf q}}
+ \dot\tau\,L = \dot{\Lambda} \,,
\end{equation}
for $\Lambda = \Lambda({\bf q},\dot{\bf q},t)$ a function of the
indicated arguments. The associated Noether invariant is
\begin{equation}
\label{eqq27} J = \tau\left(\dot{\bf q}\cdot\frac{\partial
L}{\partial\dot{\bf q}} - L\right) - {\bf\eta}\cdot\frac{\partial
L}{\partial\dot{\bf q}} + \Lambda \,.
\end{equation}

The Noether condition (\ref{eq26}) may be used to determine
simultaneously both the invariant and the action symmetry. However,
if one invariant is known, we can always obtain an associated Noether
symmetry, by use of the converse of Noether's theorem \cite{Sarlet}.
This usually leads to a symmetry of a dynamical character. In the
present case, the Ermakov invariant is available, and the procedure
may help to understand precisely its symmetry origin.  Hence, we
briefly review the converse of Noether's theorem, and apply it to our
Lagrangian Ermakov system.

The converse of Noether's theorem \cite{Sarlet}, states that if
$I({\bf q},\dot{\bf q},t)$ is an invariant for a Lagrangian dynamical
system with regular Lagrangian $L({\bf q},\dot{\bf q},t)$, then it
has the Noether symmetry (using component notation and summation
convention)
\begin{eqnarray}
\label{e1} \bar{q^i} &=& q^i +
\varepsilon\left(- g^{ij}\frac{\partial I}{\partial\dot q^j} +
\tau\dot q^i\right) \,,\\
\label{e2} \bar{t} &=& t + \varepsilon\tau
\,,
\end{eqnarray}
where $\tau = \tau({\bf q},\dot{\bf q},t)$ is an arbitrary function
and $g^{ij}$ is the inverse of the Hessian matrix of the Lagrangian,
\begin{equation} g^{ij} =
\left(\frac{\partial^{2}L}{\partial\dot{q_i}\partial\dot{q_j}}\right)^{-
1} \,.
\end{equation}
Notice that the condition $\kappa \neq 0$ assures the existence of
the inverse of the Hessian matrix.

The converse of Noether's theorem, as stated above, shows that all
invariants for regular Lagrangian systems are derivable from a
Noether symmetry which, generally, has non--point character. Hence, for
regular Lagrangian Ermakov systems, the Ermakov invariant may be
viewed as the result of a Noether symmetry. In the present case,
using the Ermakov invariant (\ref{eq3}), the Lagrangian (\ref{eq4})
and equations (\ref{e1}--\ref{e2}), we find the following Noether
symmetry associated to the Ermakov invariant $I$,
\begin{equation}
\label{eq3000} \bar{t} = t + \varepsilon\tau
\,,\qquad \bar{R} = R + \varepsilon\tau\dot{R} \,,\qquad
\bar{\theta} = \theta + \varepsilon\,(- R^{2}\dot\theta/\kappa +
\tau\dot\theta) \,,
\end{equation}
where the transformation is expressed in terms of the coordinates
$(R,\theta,t)$, and $\tau$ is a completely arbitrary function of time
and the dynamical variables.  The dynamical symmetry (\ref{eq3000})
gives directly the Ermakov invariant through Noether's theorem.
Notice that it is not possible to use
the freedom of choice of $\tau$ to reduce (\ref{eq3000}) to a point
symmetry and we necessarily have a dynamical symmetry. Consequently
the original studies \cite{RR2}--\cite{Kaushal2} that considered
only point symmetries in a one--dimensional Lagrangian formulation,
were not apt to explain the Ermakov invariant in terms of Noether
symmetries.  The existence of a (two--dimensional) Lagrangian
formalism pointed out in this work constitutes an essential
ingredient. It is important to note, however, that not all Ermakov
systems are Lagrangian and, hence the validity of the result
is still restricted.

Noether's theorem may be useful also in the search for second
invariants for Lagrangian Ermakov systems. With this goal in mind, we
restrict the treatment to point transformations. In fact, a detailed
calculation shows that allowing for dynamical symmetries depending at
most on velocities does not gives additional new results, except for the
derivation of the symmetry (\ref{eq3000}). But this symmetry is more
easily derived by direct use of the converse of Noether's theorem.
However, it is still possible that dynamical Noether symmetries
depending on higher derivatives may yield new invariants for
Lagrangian Ermakov systems.

Let us look for the existence of a second invariant for Lagrangian
Ermakov systems. For this purpose, Cartesian coordinates $(x,y)$ are
more appropriate then $(R,\theta)$ in the calculation of Noether
symmetries. The Lagrangian (\ref{eq4}), with the corresponding potential
(\ref{eq15}), becomes
\begin{equation}
\label{eq23} L = \frac{1}{2}(A\dot{x}^2 + 2B\dot{x}\dot{y} +
C\dot{y}^2) - \frac{\kappa}{R^2}\left(\int^{y/x}f(\lambda)d\lambda +
\int^{x/y}g(\lambda)d\lambda\right) -
\bar{V}(R,t) \,,
\end{equation}
where $R = R(x,y)$ is given by (\ref{eq16}).

Not  all Lagrangians of type (\ref{eq23}) possess a Noether point
symmetry. We, therefore, will search  for the least restrictive
conditions on functions $f$, $g$, $\bar{V}$ and on
the parameters $A$, $B$ and $C$, compatible with the existence of
Noether point symmetries. In such cases  the Noether theorem yields a second
invariant, independent of the Ermakov invariant.

The calculation of Noether point symmetries for a given Lagrangian is
a well known procedure \cite{Haas1} and the main steps of this
process can be summarized as follows. Inserting $L$ in the Noether
criterion (\ref{eq26}) results in a polynomial in the velocity
components, which must be identically zero.  The coefficients of the
cubic terms imply that
\begin{equation}
\label{eq27} \tau = \rho^{2}(t) \,,
\end{equation}
for any arbitrary function $\rho(t)$ of time only. This information and the
coefficients of the quadratic terms lead to
\begin{eqnarray}
\label{eq28} \eta_1 &=& \rho\dot\rho\,x - W(t)(Cy + Bx) +
a_{1}(t) \,,\\ \label{eq29} \eta_2 &=& \rho\dot\rho\,y +
W(t)(Ax + By) + a_{2}(t) \,,
\end{eqnarray}
where $a_{1}$, $a_2$ and $W$ are arbitrary functions of time.

From the coefficients of the terms linear in $\dot{x}$ and
$\dot{y}$, we get
\begin{eqnarray}
\label{eq30} \partial\Lambda/\partial x &=&
A\partial\eta_{1}/\partial t + B\partial\eta_{2}/\partial t \,,\\
\label{eq31}
\partial\Lambda/\partial y &=& C\partial\eta_{2}/\partial t +
B\partial\eta_{1}/\partial t \,.
\end{eqnarray}
These last two equations will have a solution $\Lambda$ provided
that $\partial^{2}\Lambda/\partial{x}\partial{y} =
\partial^{2}\Lambda/\partial{y}\partial{x}$. This implies, after
a simple calculation involving $\eta_1$ and $\eta_2$ given by
(\ref{eq28}--\ref{eq29}), that
\begin{equation}
\label{eqq31} \dot W = 0 \,.
\end{equation}
The corresponding solution for (\ref{eq30}--\ref{eq31}) is
\begin{equation}
\label{eq32} \Lambda = \Lambda_{0}(t) + \frac{1}{2}(\rho\ddot\rho
+ \dot\rho^2)R^2 + \dot{a}_{1}(Ax + By) + \dot{a}_{2}(Bx + Cy) \,,
\end{equation}
where $\Lambda_{0}(t)$ depends on time only. The remaining term,
independent of velocity components in the Noether
symmetry condition, yields a first order linear partial
differential equation for the potential,
\begin{equation}
\label{eq33} \rho^{2}\frac{\partial V}{\partial t} +
{\bf\eta}\cdot\frac{\partial V}{\partial{\bf q}} = -
2\rho\dot\rho\,V - \frac{\partial\Lambda}{\partial t} \,.
\end{equation}
The potential $V$ must comply with the Ermakov form (\ref{eq15}), and
hence, is not completely arbitrary. Inserting $\eta_1$, $\eta_2$,
$\Lambda$ and the Ermakov form for $V$, we get an equation for
$\bar{V}$,
$$
\rho^{2}\frac{\partial\bar{V}}{\partial t} +
\rho\dot\rho\frac{\partial\bar{V}}{\partial R} +
2\rho\dot\rho\bar{V} = $$
\begin{equation}
- \dot{\Lambda}_0 - \frac{1}{2}(\rho{\buildrel\cdots\over{\rho}} +
3\dot\rho\ddot\rho)R^2  - \ddot{a_1}(A\,x + B\,y)
- \ddot{a_2}(B\,x + C\,y) +
\end{equation}
$$
\frac{2\kappa}{R^4}\left(a_{1}(A\,x + B\,y) + a_{2}(B\,x +
C\,y)\right)\left(\int^{y/x}\!\!\!f(\lambda)d\lambda +
\int^{x/y}\!\!\!g(\lambda)d\lambda\right) + $$ $$
\kappa\left(\frac{a_{1}y - a_{2}x}{R^2} -
W\right)\left(\frac{f}{x^2} - \frac{g}{y^2}\right) \,.$$
Requiring $f$ and $g$ to remain arbitrary implies that the $f$ and $g$
dependent parts of the last equation vanish. This is obtained by
choosing
\begin{equation}
\label{eq34}
W = a_1 = a_2 = 0 \,.
\end{equation}
Consequently,
\begin{equation}
\label{eq35} \rho^{2}\frac{\partial\bar{V}}{\partial t} +
\rho\dot\rho\frac{\partial\bar{V}}{\partial R} +
2\rho\dot\rho\bar{V} = - \dot{\Lambda}_0 -
\frac{1}{2}(\rho{\buildrel\cdots\over{\rho}} +
3\dot\rho\ddot\rho)R^2 \,.
\end{equation}
The general solution of (\ref{eq35}) is
\begin{equation}
\label{eq36} \bar{V}(R,t) = - \frac{\Lambda_0}{\rho^2} -
\frac{\ddot\rho}{2\rho}R^2 + \frac{1}{\rho^2}U(R/\rho) \,,
\end{equation}
where $U(R/\rho)$ is an arbitrary function of the indicated
argument. In fact, we can take $\Lambda_0 = 0$ without any loss of
generality since the addition of a function of time only in the
Lagrangian will not affect the equations of motion. This completes
the Noether point symmetry calculation.

We remark that (\ref{eq36}) could have been obtained from the
potential associated to the autonomous Hamiltonian Ermakov systems
in \cite{Haas2} by performing a rescaling transformation $\tilde{x}
= x/C(t)$, $\tilde{y} = y/C(t)$, $\tilde{t} = \int\,dt/C^{2}(t)$
and requiring the resulting equations to stay autonomous.
However, Noether's theorem provides a more consistent basis for the
reasons why the potential (\ref{eq36}) is integrable: integrability
here is viewed as resulting from a general symmetry consideration
and not from an {\it ad hoc} transformation. Moreover, the
calculation of Noether point symmetries provides a  hint for the
determination of the more general dynamical Noether symmetries.

Let us summarize our results so far. From Noether's symmetry
criterion applied to the class of Lagrangian Ermakov systems of
section 2, we  obtained the symmetry transformations
\begin{equation}
\label{eq37} \bar{x} = x + \varepsilon\rho\dot\rho\,x \,, \qquad
\bar{y} = y + \varepsilon\rho\dot\rho\,y \,, \qquad \bar{t} = t +
\varepsilon\rho^2 \,.
\end{equation}
The corresponding potentials can be expressed in terms of $f$, $g$
and another arbitrary function $U(R/\rho)$ as
\begin{equation}
\label{eq38} V = - \frac{\ddot\rho\,R^2}{2\rho} +
\frac{1}{\rho^2}U(R/\rho) +
\frac{\kappa}{R^2}\left(\int^{y/x}f(\lambda)d\lambda +
\int^{x/y}g(\lambda)d\lambda\right) \,.
\end{equation}
Consequently four arbitrary functions, namely $\rho(t)$, $U(R/\rho)$,
$f(y/x)$ and $g(x/y)$, remain in the system. In $(R,\theta)$
coordinates, the Ermakov invariant and the invariant arising from point
symmetry have the compact representation,
\begin{eqnarray}
\label{eq40} I &=& \frac{1}{2}R^{4}\psi^{4}(\theta)\dot\theta^2 +
\int^{\tan\theta}f(\lambda)d\lambda +
\int^{\cot\theta}g(\lambda)d\lambda \,,\\
\label{eqq41}
J &=& \frac{1}{2}(\rho\dot R - \dot\rho\,R)^2 + U(R/\rho) +
\kappa\,I\left(\rho/R\right)^2 \,.
\end{eqnarray}
%
For the sake of comparison, we writte, using configuration space variables,
the corresponding non-constant, time-dependent Hamiltonian function which
follows from (\ref{HHH}),
\begin{equation}
H = \frac{1}{2}\dot{R}^2 + \frac{\kappa\,I}{R^2} -
\frac{\ddot\rho\,R^2}{2\rho} + \frac{1}{\rho^2}U(R/\rho) \,.
\end{equation}

An important point to check is the fact that $I$ and $J$ are in
involution. This follows from the conventional definition of Poisson
brackets and the relations $\dot{x} = (Cp_x - Bp_{y})/\kappa$,
$\dot{y} = (Ap_y - Bp_{x})/\kappa$.

A simple inspection shows that level surfaces of the function $I$ are not
compact, a fact that prevents the application of Liouville's theorem.
Despite this, the exact solution for the Ermakov systems possessing
the invariants (\ref{eq40}--\ref{eqq41}) can be obtained as follows.
For given $\rho(t)$, define new variables
\begin{equation}
\label{resca}
\bar{R} = R/\rho \,,\quad T = \int{dt}/\rho^2
\end{equation}
and transform the Noether invariant (\ref{eq40}) into
\begin{equation}
J = \frac{1}{2}(d\bar{R}/dT)^2 + U(\bar{R}) +
\kappa\,I/\bar{R}^2 \,,
\end{equation}
an energy--like form. Using the energy method \cite{Arnold}, we can
find, by quadratures, $\bar{R}$ as a function of $T$, and then $R$ as
a function of $t$. The exact solution is then obtained by inserting
$R$ as a function of $t$ in the Ermakov invariant (\ref{eq40}), which
can then be interpreted as a first-order ordinary differential
equation for the solution $\theta$. Being separable, this equation is
again soluble by quadratures.

We observe that the Noether symmetry (\ref{eq37}) is of the same type
as the Lie point symmetry eventually admitted by Ermakov systems
\cite{Leach}, as expected.  Indeed,
Noether point symmetries form a subgroup of the more general Lie
point symmetry group of the same dynamical system \cite{Sarlet}. At
least in this case, however, the Noether symmetries result more
interesting because they, directly, produce constants of motion.

In spite of the reduction of the Lagrangian Ermakov systems with Noether point symmetry to quadrature, these systems are not linearisable, in general. Indeed, using
 (\ref{linearV}) and
(\ref{eq36}) we can show that the only two cases compatible with linearisation are
\begin{equation}
\label{U1}
U(\bar{R}) = a/\bar{R}^2 - b/\bar{R} \,,
\end{equation}
or
\begin{equation}
\label{U2}
U(\bar{R}) = a/\bar{R}^2 + c\bar{R}^{2}/2 \,,
\end{equation}
for $a$, $b$ and $c$ arbitrary numerical constants. Moreover, when $U$ has the form (\ref{U1}), the function $\alpha$ in (\ref{varphi}) must be choosen as $\alpha = \rho$. When $U$ has the form (\ref{U2}), $\alpha$ must be choosen as any particular solution for the equation
\begin{equation}
\ddot\alpha + \left(\frac{c}{\rho^3} - \ddot\rho\right)\frac{\alpha}{\rho} = 0 \,.
\end{equation}
In all linearisable cases, the function $\rho(t)$ is left free.

We finally compare our result with other results in the
literature. Grammaticos and Dorizzi have derived the potential
(\ref{eq38}) with $A = C = 1$, $B = 0$
and arbitrary $\rho(t)$ (see equation (60) of
\cite{Grammaticos}) in their search for two--dimensional
systems with invariants quadratic in velocity. However, they did not
recognize the Ermakov character of the system. Dhara and
Lawande, in cases A) and C) in reference  \cite{Dhara}, found
potentials that correspond to Ermakov systems or to Noether point
symmetry. However, the possible intersection of the two classes of
potentials, yielding Ermakov systems with Noether point symmetry, was
not explored at all. Leach, Lewis and Sarlet \cite{LLS} found a class
of potentials with an invariant quadratic in velocity.  Within this
class, there exists an Ermakov subclass, a fact that was not pointed out
precisely. The Ermakov structure in these systems was hidden due to
the non usual character of the frequency function, depending on
dynamical variables. Finally, the second invariant found by Goedert
\cite{Goedert} is a Noether invariant which can be obtained in the
present formalism, by choosing $A = C = 0$, $B = 1$
and $\rho = 1$.

\section{Conclusion}

In this paper we identified a subclass of  Ermakov systems with two
interesting properties, namely, of being Lagrangian and of
possessing Noether symmetries of both point and dynamical character.
We  also show, by the converse of
Noether's theorem, how the Ermakov invariant follows from a Noether
symmetry of dynamical character.  This is an alternative criterion
to characterize and classify the Ermakov systems which should be
considered parallel with the existence of Lie point symmetries
\cite{Haas3,Leach} or of special
structure of the solutions on the complex plane \cite{Athorne3}.
Moreover, the class of Lagrangian Ermakov systems with
Noether point symmetries presented in this paper may possibly be
expanded further by relaxing our starting point in terms of the ansatz
(\ref{eq4}) for the Lagrangian.

\noindent\bigskip\newpage
\leftline{\bf Acknowledgments} \smallskip
This work was partially supported by Funda\c{c}\~ao de Amparo \`a
Pesquisa do Estado do Rio Grande do Sul (Fapergs) and Conselho
Nacional de Desenvolvimento Cient\'{\i}fico e Tecnol\'ogico (CNPq).
The authors also acknowledge the Instituto de F\'{\i}sica (IF-UFRGS)
for continued support of their research projects.

\end{document}